\documentstyle[aps,epsfig,prl,floats]{revtex}
\newcommand{\be}{\begin{equation}}
\newcommand{\ee}{\end{equation}}
\newcommand{\bea}{\begin{eqnarray}}
\newcommand{\eea}{\end{eqnarray}}

\tightenlines
\begin{document}
\draft

\title{Deconfinement Transition and Bound States 
in Frustrated Heisenberg Chains: Regimes of
Forced and Spontaneous Dimerization}

\author{Weihong Zheng\cite{zwh} and Chris J.~Hamer \cite{cjh}}
\address{
School of Physics, University of New South Wales, Sydney NSW 2052, Australia}
\author{Rajiv R.~P.~Singh}
\address{
Department of Physics, University of California, Davis, CA 95616}
\author{Simon Trebst \cite{Bell} and Hartmut Monien}
\address{
Physikalisches Institut, Universit\"at Bonn, Nu\ss allee 12, 53115 Bonn, 
Germany}
                  
\date{\today}

\maketitle

\begin{abstract}
We use recently developed strong-coupling expansion methods to
study the two-particle spectra for the frustrated alternating
Heisenberg model, consisting of an alternating nearest neighbor
antiferromagnetic exchange and a uniform second neighbor
antiferromagnetic exchange. Starting from the limit of weakly coupled dimers,
we develop high order series expansions for the effective Hamiltonian
in the two-particle subspace. In the limit of a strong
applied dimerization, we calculate accurately various properties of
singlet and triplet bound states and quintet
antibound states. We also develop series expansions for bound state
energies in various sectors, which can be extrapolated using
standard methods to cases where the external
bond-alternation goes to zero. We study the properties of singlet and
triplet bound states in the latter limit and suggest a crucial role
for the bound states in the unbinding of triplets and deconfinement of spin-half
excitations.
\end{abstract}

\section{INTRODUCTION}

        In recent years the subject of fractional excitations, or
excitations with fractional quantum  numbers compared to
the non-interacting limit, has attracted considerable attention.
Many experiments in high temperature superconducting
materials have been interpreted in these terms, and
many theories of high temperature superconductors are built
around such fractional excitations.
However, fractional excitations  in a closed
system can only arise in groups that have the full quantum numbers of the
non-interacting limit. Thus, their unambiguous identification in
numerical calculations remains difficult.

In this work we are interested in
studying models, where as parameters in the Hamiltonian
are varied, one goes from a phase where the excitations
have normal quantum numbers to one where they have fractional
quantum numbers. We would like to develop series expansion 
methods by which such transitions can be studied
and the onset of fractional
excitations can be demonstrated. Although we work with 1D
models, the basic methods we develop can be applied in
higher dimensions as well.
 
The best known example of a fractional excitation is a
spinon in the spin-half antiferromagnetic Heisenberg chain \cite{Faddev:81}.
It is well-known that the low-lying excitations
of this Bethe-ansatz solvable model consist of a two-spinon continuum.
Another simple example of a spin-half excitation is a soliton in
the Majumdar-Ghosh model\cite{MG}. It is a domain-wall which interpolates
between the two dimerized ground states of the model. 
The low-lying excitations, for a closed system, in this case also consist of a 
soliton-antisoliton continuum \cite{Shastry:81}.

Adding a bond-alternation to the exchange constants of the model
leads to confinement of the spin-half excitations. The elementary
excitations now become triplets and the spinons or solitons are
bound into pairs. In the unfrustrated case, the role of
bond-alternation has been studied by mapping onto a massive
Thirring model\cite{Haldane:82}. The frustrated case, where the ground-states
are spontaneously dimerized, has been of considerable recent theoretical
interest due to its relevance to spin-Peierls systems such as
CuGeO$_3$. An explicit bond-alternating term in the Hamiltonian
can be motivated as a mean-field representation of the inter-chain
elastic couplings\cite{Affleck:97}. Uhrig {\it et al.}\cite{Uhrig:99} and
Affleck and collaborators \cite{sor98} have
studied the confinement transition for the soliton-antisoliton pairs
when such a term is added to the Hamiltonian.

Here, we approach these transitions from the opposite direction.
We study these systems in a strong-coupling perturbation theory,
which begins with the limit of decoupled spin-dimers and treats
the inter-dimer couplings as a perturbation. In the limit of weakly
coupled dimers, the elementary excitations are triplets, which are
weakly dispersive. In this limit our strong coupling theory is
highly accurate and we can find all details of various two-particle
bound (and antibound) states. The overall 2-particle spectrum is
much richer than that obtained in previous studies.
Several singlet and triplet bound states and quintet antibound
states are found. The number of bound states depends on the coupling constants
as well as the wavevector.
We study the binding energy 
and the coherence length associated with the bound states.
We also study the singularity at the critical wavevectors
where the binding energy goes to zero and the state merges into the continuum.

Using high order series expansions
and extrapolation methods we also study the uniform limit,
where the bond-alternation term in the Hamiltonian goes to zero.
Thus we approach the limit where the triplet excitations break up
and spin-half excitations become deconfined. Series expansion
results show that as the bond-alternation term goes to zero in the
Hamiltonian, the spectral weights associated with triplet quasiparticles
go to zero and the lowest lying singlet and triplet excitations
become degenerate.
These phenomena provide a remarkably clear and simple confirmation of
the existence of free spin-half excitations in this limit.

In the frustrated system,
the reorganization of the many-body spectra as the system
undergoes the deconfinement transition presents an interesting
puzzle. Since there is a gap, $\Delta$, to triplet excitations,
the two-triplet continuum begins at $2\Delta$. Thus, in the confined phase,
this continuum is separated
from the elementary triplet by a second gap. However, when the
spin-half excitations are liberated, the resulting low energy
spectrum consists of a soliton-antisoliton continuum, which does
not have such a second gap. The consistency of the two pictures
requires that in the confined phase, between the elementary triplet and the
two-triplet continuum there must be a large number of states,
which upon deconfinement turn into the continuum. On general
grounds, these states must include $2$, $3$, $4$, $\ldots$,
triplet bound states, i.e. states involving an arbitary number of triplets,
which must correspond to a soliton-anitisoliton pair with arbitrary separation.
We discuss insights from studies of two-particle bound states on
this issue.

Another interesting puzzle lies in the spectrum of the
Majumdar-Ghosh model ($\delta=0$, $\alpha=1/2$) near $k=\pi/2$.
For both $S=0$ and $S=1$, previous studies\cite{Shastry:81,oleg99,sor99}
 have emphasized a 
bound state below the
soliton-antisoliton ($s-\bar s$) continuum. To our knowledge,
it has not been noticed that the two triplet continuum
(the $s-\bar s-s-\bar s$ continuum) falls below the 
soliton-antisoliton continuum around this wavevector.
In our numerical study, we find that
except for a tiny region very near $k=\pi/2$, the latter
continuum also falls below the bound states. This raises
questions about the stability of the bound states away from $k=\pi/2$. 

Another puzzle in our studies is how the energy levels might cross each other.
On general grounds, one might expect that the levels for
$n$-particles with varying $n$ cross each other as the parameters
are varied. For small values of the perturbation parameter, the
energies are arranged in order of increasing $n$. However, as
one approaches $\lambda=1$, low energy states from each $n$-sector
may appear even below the two-particle continuum. Since $n$ is not
a good quantum number, it is not clear how this will reflect itself
in our perturbation theory. This 
deserves further attention.

The organization of the paper is as follows. In section II,
we describe the Hamiltonian studied and the various parametrizations
used. In section III, we study the regime of forced or
externally imposed dimerization. This is a regime where our series
expansions are convergent and we present spectra, binding energies,
coherence lengths etc. in great detail. In section IV, we consider
the regime of spontaneous dimerization, which requires the use of
 series extrapolation methods. In section V, we present discussions
and conclusions.

\section{Hamiltonian}

We wish to study the alternating Heisenberg chain with frustration
\cite{Shastry:81,sor98,Uhrig:96,Rajiv:99,Bouzerar:98,Kotov:99b,Barnes:99,oleg99},
\begin{equation}
H =  \sum_i {[(1+(-1)^i\delta) {\bf S}_i\cdot {\bf S}_{i+1}
               + \alpha {\bf S}_i\cdot {\bf S}_{i+2}]} \;,
\label{H:chain}
\end{equation}
where the ${\bf S}_i$ are spin-$\frac{1}{2}$ operators at site $i$,
$\alpha$ parameterizes a next-nearest neighbor coupling and $\delta$ is
the 
alternating dimerization. We rewrite the Hamiltonian as
\begin{equation}
H= (1 + \delta) \sum_i {[{\bf S}_{2i}\cdot {\bf S}_{2i+1}
+\lambda ( {\bf S}_{2i}\cdot {\bf S}_{2i-1}
               + y {\bf S}_i\cdot {\bf S}_{i+2} ) ]} \;,
\label{H:chain2}
\end{equation}

The parameter space ($\delta$, $\alpha$) is equivalent to the
parameter space ($\lambda$, $y$) with $\lambda\equiv (1-\delta)/(1+\delta)$
and $y\equiv \alpha/(1-\delta)$. The latter parametrization
makes explicit that for $\lambda=0$, the model consists of
decoupled dimers: we take this to be our unperturbed
Hamiltonian $H_0$. The rest of the Hamiltonian can be treated as
a perturbation, and we can expand various physical quantities
in powers of $\lambda$. The formalism for studying n-particle
sectors in perturbation theory is discussed in detail in
a companion paper \cite{longpaper}.

The series expansions for the ground state energy and triplet excitation 
spectrum
have previously  been computed \cite{Rajiv:99} up to order 23.
The two-particle excitations have been discussed using a leading order 
Brueckner ansatz calculation\cite{Kotov:99b}, a second order
series expansion\cite{Barnes:99} and an RPA study\cite{Uhrig:96}.
With our new technique, we perform high-order series expansions in powers
of $\lambda$ for fixed values of $y$ \cite{wwwsite}.
As discussed in a companion paper \cite{longpaper}, 
we first calculate an effective Hamiltonian in the two-particle sector
\begin{equation}
E_{2}({\bf i,j;k,l})=\langle {\bf k,l} | H^{\rm eff} | {\bf i,j} \rangle \;,
\end{equation}
and then calculate the irreducible two-particle matrix element
\begin{eqnarray}
\Delta_{2}({\bf i,j;k,l}) &  = & E_{2}({\bf i,j;k,l}) -E_{0}(\delta_{{\bf
i,k}}\delta_{{\bf j,l}} + \delta_{{\bf i,l}}\delta_{{\bf j,k}}) -
\Delta_{1}({\bf i,k})\delta_{{\bf j,l}} - \Delta_{1}({\bf i,l})
\delta_{{\bf j,k}} \nonumber\\
 & & - \Delta_{1}({\bf j,k}) \delta_{{\bf i,l}} -
\Delta_{1}({\bf j,l}) \delta_{{\bf i,k}} \;,
\end{eqnarray}
where $\delta$ refers to a Kronecker delta function and
$\Delta_{1}$ is the one-particle irreducible matrix element.

Here we will only concentrate on the expansions for the following two
lines in the parameter space: 
(i) $\alpha=0$, corresponding to nearest neighbor
interaction only,
and (ii) $\alpha=(1- \delta)/2$, which is a special line in
the parameter space where the ground states are known exactly,
also known as the Shastry-Sutherland line. The model
at $\delta=0$  $(\lambda=1)$ is the uniform Heisenberg chain in case (i)
and the Majumdar-Ghosh model in case (ii).

\section{Bound States with Forced Dimerization}

In this section we study the small-$\lambda$ regime, where our strong coupling expansions
are convergent. Thus a simple truncation of the 
relevant power series expansions
leads to highly accurate results. We discuss the number of bound
states with different quantum numbers as well as their
binding energies, the range of k-values where
the bound states exist, the coherence length associated with
the bound-pair as well as the singularities at the critical
wavevector, where the binding energy goes to zero and the bound
state merges into the continuum. We first consider the model
with $\alpha=0$ and then $\alpha=(1-\delta)/2$.

\subsubsection{Case $\alpha=0$}
For the case  without the second neighbor interaction ($\alpha=0$),
the series for the irreducible two-particle matrix element $\Delta_2$ have been computed up to order 7 for singlet
bound states, and to order 11 for triplet and quintet states\cite{wwwsite}. 
The reason why the singlet series is computed to only 7th order compared
to 11th order for the triplet and quintet states is that the
singlet has the same quantum numbers as the ground state. Thus
a much more elaborate orthogonalization method is required to implement the
cluster expansion \cite{longpaper}.

In this model, we find two singlet ($S_1$ and $S_2$) and two triplet
($T_1$ and $T_2$) 
bound states below the two-particle continuum, and two quintet antibound states
($Q_1$ and $Q_2$)
above the continuum.  
The existence of the second pair of bound states has not been reported
by previous calculations, most likely due to a limited precision or a
general
incapability to deal with multiple bound states.
The series for their energies (and also the lower edge and upper edge of the 
continuum) at band maximum  $k=\pi/2$ are given in Table \ref{tabJ12d_0}. 
Note that there are some discrepancies for the energy of the lowest singlet bound state 
with the previous second order calculations\cite{Kotov:99b}.
Our second order results agree with the series results 
 of Barnes {\it et al.}\cite{Barnes:99}, but disagree with the
 results of the Brueckner ansatz calculation\cite{Kotov:99b}.
Although the Brueckner ansatz is an expansion of the self-energy in terms of the 
density of excitations,  it normally can recover the first few order  of the 
series expansion in $\lambda$ exactly:  it does give the correct second order result 
for the triplet bound state.\cite{Kotov:99b}

In the limit $\lambda\to 0$, the formation of bound/antibound states $S_1$, $T_1$ 
and $Q_1$ is well known, due simply to the interaction  
of two triplets on neighboring sites, 
and the wave function for $S_1$, for example, is
\be
\vert \psi_{S_1} (k) \rangle = {1\over \sqrt{3N}} \sum_{j} e^{2ik(j+1/2)}  
[t_{1}^{\dag} (j) t_{-1}^{\dag} (j+1)  + t_{-1}^{\dag} (j) t_{1}^{\dag} (j+1)
- t_{0}^{\dag} (j) t_{0}^{\dag} (j+1) ] \vert 0 > \;,
\ee
where $\vert 0\rangle$ is the ground state at $\lambda=0$ 
consisting of nonoverlapping spin singlets on each dimer, and $t_{\alpha}^{\dag} (j)$
is a triplet creation operator which excites the singlet at $j$-th dimer into a triplet 
state with $S_z=\alpha$,
($\alpha=-1,0,1$).

It is interesting also to look at the structure of these 
new bound states $S_2$, $T_2$ and $Q_2$ in this limit.
To compute their wavefunctions, one needs to diagonalize the second
order effective Hamiltonian in the two-particle sector, 
which can be reduced to an infinite dimensional symmetric tridiagonal
matrix.
Our calculations show that in this limit,  
the bound/antibound states $S_2$, $T_2$ and $Q_2$ only exist at $k=\pi/2$, and 
their wave functions are (we take $S_z=0$ as example)
\bea
\vert \psi_{S_2} (k) \rangle =&& {1\over \sqrt{3N}} \sum_{j} e^{2ikj} \sum_{n=1}^{\infty} f_n 
[t_{1}^{\dag} (j-n) t_{-1}^{\dag} (j+n)\nonumber \\
&&  + t_{-1}^{\dag} (j-n) t_{1}^{\dag} (j+n)
- t_{0}^{\dag} (j-n) t_{0}^{\dag} (j+n) ] \vert 0 > \;, \\
\vert \psi_{T_2} (k) \rangle =&& {1\over \sqrt{2N}} \sum_{j} e^{2ikj} \sum_{n=1}^{\infty} f_n
 [ t_{1}^{\dag} (j-n) t_{-1}^{\dag} (j+n) 
- t_{-1}^{\dag} (j-n) t_{1}^{\dag} (j+n) ] \vert 0 > \;, \\
\vert \psi_{Q_2} (k) \rangle =&& {1\over \sqrt{6N}} \sum_{j} e^{2ikj} \sum_{n=1}^{\infty} f_n
[t_{1}^{\dag} (j-n) t_{-1}^{\dag} (j+n)\nonumber \\
&&  + t_{-1}^{\dag} (j-n) t_{1}^{\dag} (j+n)
+ 2 t_{0}^{\dag} (j-n) t_{0}^{\dag} (j+n) ]  \vert 0 > \;,
\eea
where the amplitude $f_n$ for two triplets sitting at $j-n$ and $j+n$  
(i.e. separated by distance $2n$) is
\be
f_n = \left\{ 
\begin{array}{ll}
 -\sqrt{15} (-4)^{-n} & \mbox{for $S_2$} \\
 -\sqrt{3} (-2)^{-n} & \mbox{for $T_2$} \\ 
  \sqrt{3} 2^{-n} & \mbox{for $Q_2$} \;.
\end{array}
\right.
\ee
Thus the formation of these new bound/antibound states is due to an effective 
interaction between triplets separated by an odd number of singlet dimers.
It appears that $S_1$, $T_1$ and $Q_1$ are fully ``localized" states in
this limit, with wavefunctions extending only across a single pair of dimers,
whereas  the states $S_2$, $T_2$ and $Q_2$ are ``extended", with the tail of the
wavefunctions decreasing exponentially with distance. 
%
%
Our calculations show that at finite $\lambda$ and for the particular
case $k=\pi/2$, the wavefunctions for $S_2$,  $T_2$ and $Q_2$ still only involve
 triplets separated by an odd number of singlet dimers, while 
the wavefunctions for $S_1$,  $T_1$ and $Q_1$ only involve triplets 
separated by an even number of singlet dimers.
Thus the spectrum splits into two decoupled sectors at $k=\pi/2$.
It would be interesting to explore the dynamics 
behind this phenomenon  in more depth.

With the wavefunction, one can also compute the coherence length $L$
defined by
\be
L = {\sum_{d=1}^{\infty} d f_d^2\over \sum_{d=1}^{\infty} f_d^2}
\ee
where $f_d$ is the amplitude for two triplets separated by distance $d$.
Note that the coherence length $L$ defined here is measured in units of $2a$,
where $a$ is the lattice spacing. At the limit $\lambda\to 0$,
$L=1$ for $S_1$ and $T_1$ as expected,
while for $S_2$ and $T_2$, $L$ is 32/15 and 8/3, respectively.

\begin{figure}
 \begin{center} 
 \vskip -0.5cm
 \epsfig{file=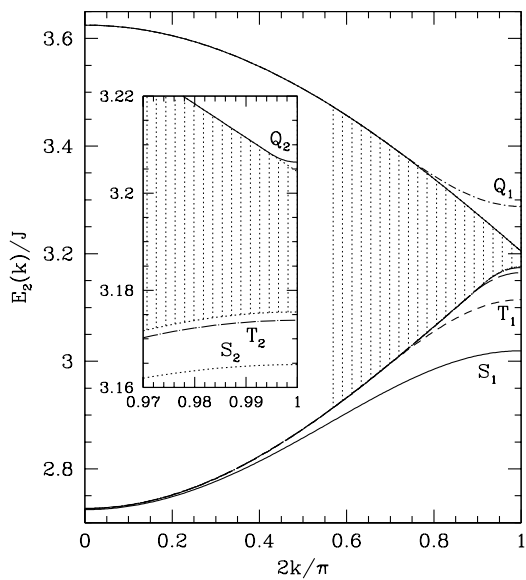, height=10cm}
 \vskip -1.2cm
 \caption[]
         {The excitation spectrum of the $J_1-J_2-\delta$ chain with
          $\delta=0.6$ and $\alpha=0$. Beside the two-particle
          continuum (gray shaded), there are
          two singlet bound states ($S_1$ and $S_2$) and
          two triplet bound states ($T_1$ and $T_2$) below the continuum, and two
          quintet antibound states ($Q_1$ and $Q_2$) above the continuum.
          The inset enlarges the region near 
          $k=\pi/2$ so we can see $S_2$, $T_2$ and $Q_2$ below/above the continuum.}
 \label{figJ12d_0_mk} 
 \end{center}
\end{figure}

\begin{figure}
 \begin{center} 
 \vskip -0.5cm
 \epsfig{file=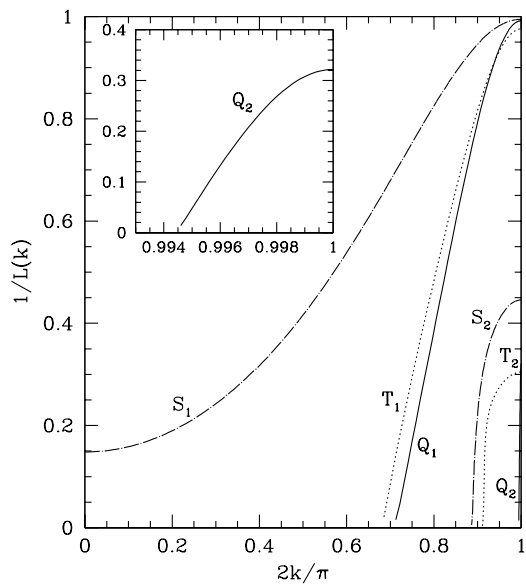, height=10cm}
 \vskip -1.2cm
 \caption[]
          {The inverse of the coherence length $1/L$ 
          versus momentum $k$ for two singlet ($S_1$ and $S_2$), 
          two triplet ($T_1$ and $T_2$) and two quintet ($Q_1$ and $Q_2$) 
          bound/antibound states of the $J_1-J_2-\delta$ chain 
          with $\delta=0.6$ and $\alpha=0$. The inset enlarges the region near 
          $k=\pi/2$.}
 \label{figJ12d_0_CL_dp6} 
 \end{center}
\end{figure}

\begin{figure}
 \begin{center} 
 \vskip -0.5cm
 \epsfig{file=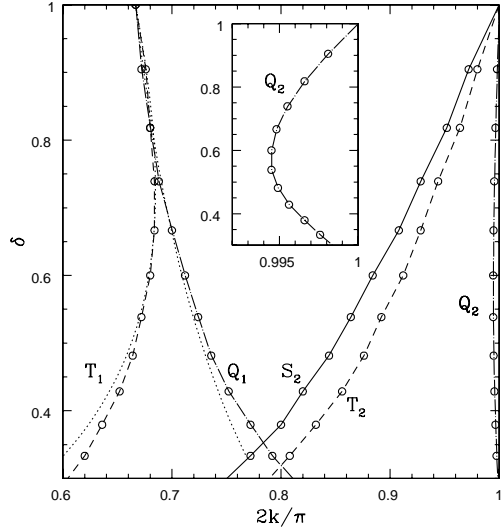, height=10cm}
  \vskip -1.2cm
 \caption[]
         {The critical $\delta$ versus momentum $k$ for singlet, triplet  and quintet
          bound/antibound states of the $J_1-J_2-\delta$ chain with $\alpha=0$. 
         The dotted lines are the
         results of Eq. (\ref{eqJ12d0Kc}). }
 \label{figJ12d_0_pd} 
 \end{center}   
\end{figure}

The two-particle excitation spectrum and the inverse of the coherence length $1/L$ 
versus momentum $k$ for 
a rather large dimerization $\delta=0.6$ are shown 
in Figs. \ref{figJ12d_0_mk} and \ref{figJ12d_0_CL_dp6}.
One can see that the singlet bound state $S_1$ exists for the whole range 
of momenta (its coherence length $L$ is finite also for the whole range 
of momenta although the coherence length at $k=0$ is very large, about 6.7706), while other
bound/antibound states exist only in a limited range of momenta $k>k_c$. The
``critical momentum" $k_c$ for a given bound state can be defined
 by the inverse of the coherence length $1/L$ tending to zero
or by vanishing binding energy. 
Technically, the first approach may give more reliable results on $k_c$.
The results for
the critical momenta $k_c$ versus $\delta$ 
are given in Fig. \ref{figJ12d_0_pd}, where we can see that 
in the limit $\delta\to 1$, the $S_2$, $T_2$ and $Q_2$ states exist only at $k=\pi/2$.
We can also 
get the first few terms in the series expansion for
 $k_c$  for $T_1$ and $Q_1$ states.
Up to order $\lambda^3$, the dispersions for the
 $T_1$ and $Q_1$ states are
\begin{mathletters}
\label{eqJ12d0}
\bea
E_{T_1}/J (1+\delta) &=& 
  2 - {\frac{3\,\lambda}{4}} - {\frac{9\,{\lambda^2}}{32}} + {\frac{89\,{\lambda^3}}{128}} + 
   \left( - {\frac{\lambda}{2}} - {\frac{3\,{\lambda^2}}{16}} + 
      {\frac{105\,{\lambda^3}}{128}} \right) \,\cos (2k) \nonumber \\
      && + 
   \left( {\frac{{\lambda^2}}{4}} - {\frac{3\,{\lambda^3}}{16}} \right) \,\cos (4\,k) - 
   {\frac{37\,{\lambda^3}\,\cos (6\,k)}{128}} + O(\lambda^4) \;, \\
E_{Q_1}/J (1+\delta) &=& 2 + {\frac{3\,\lambda}{4}} - {\frac{{\lambda^2}}{32}} + {\frac{21\,{\lambda^3}}{128}} + 
   \left( {\frac{\lambda}{2}} - {\frac{3\,{\lambda^2}}{16}} + {\frac{37\,{\lambda^3}}{128}}
       \right) \,\cos (2k) \nonumber \\
       && + \left( -{\frac{{\lambda^2}}{4}} + 
      {\frac{13\,{\lambda^3}}{64}} \right) \,\cos (4\,k) + 
   {\frac{23\,{\lambda^3}\,\cos (6\,k)}{128}}  + O(\lambda^4) \;.
\eea
\end{mathletters}
With this and the series for the 1-particle triplet excitation spectrum, one can get
$k_c$ as
\be
2k_c = \left\{ \begin{array}{ll}  
2 \pi/3 + 5 \lambda/(4 \sqrt{3}) - 757 \lambda^2/(192 \sqrt{3}) + O(\lambda^3) & \mbox{for $T_1$} \\
2 \pi/3 + \sqrt{3} \lambda/4 + 15 \sqrt{3}\lambda^2/64 + O(\lambda^3)  & \mbox{for $Q_1$}
\end{array}
\right.
\label{eqJ12d0Kc}
\ee
and in the limit $k\to k_c$, the behaviour of the binding energy is
\bea
E_b/J\lambda &=& 4(k-k_c)^2 [3/16 - 11 \lambda/128+ 591\lambda^2/512 
    + O(\lambda^3) ] \nonumber \\
&& + 8(k-k_c)^3 [ \sqrt{3}/32 + 113 x/(256\sqrt{3}) 
  - 25 \lambda^2/(768 \sqrt{3}) + O(\lambda^3) ] 
    +  O[(k-k_c)^4]
\eea
for $T_1$, and
\bea
E_b/J\lambda &=&  
4(k-k_c)^2 [ {\frac{3}{16}} - {\frac{21\,\lambda}{128}} - {\frac{39\,{\lambda^2}}{256}}
 + O(\lambda^3)] \nonumber \\
&&+ 8 \sqrt{3} (k-k_c)^3 [ {\frac{1}{32}} + {\frac{37\,\lambda}{256}} 
- {\frac{37\,{\lambda^2}}{1024}} + O(\lambda^3) ]  + O[(k-k_c)^4]
\eea  
for $Q_1$.
Here one can see that the ``critical index" for $E_b$ in the limit $k\to k_c$ is 2, 
independent of the order of expansion, so one expects that this is {\it exact}.
The results of Eq. (\ref{eqJ12d0Kc}) are also shown in Fig. \ref{figJ12d_0_pd}.
We can see that in the limit $\delta\to 1$, $k_c=\pi/3$ for $T_1$ and $Q_1$,
and as $\delta$ decreases, $k_c$ for $Q_1$ increases, while $k_c$ for $T_1$
firstly increases, then decreases.

\begin{figure}
 \begin{center} 
 \vskip -0.5cm
 \epsfig{file=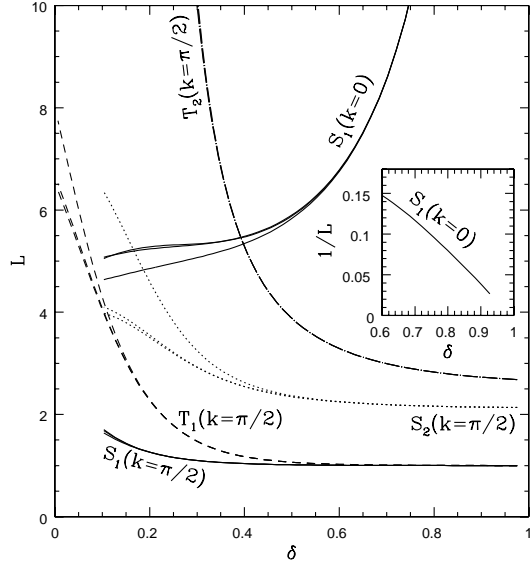, height=10cm}
 \vskip -1.2cm
 \caption[]
          {The  coherence length $L$ 
          versus $\delta$ for two singlet ($S_1$ and $S_2$) and
          two triplet ($T_1$ and $T_2$) 
          bound states of the $J_1-J_2-\delta$ chain 
          at $k=0$, $\pi/2$ and $\alpha=0$. The inset plots $1/L$ versus $\delta$
          for $S_1$ at $k=0$. The results of the three highest orders are plotted. }
 \label{figJ12d_0_CL_k0pi} 
 \end{center}
\end{figure}

Actually $S_1$ does not always exist in the whole range 
of momenta: it  does not exist at $k=0$ when $\delta\to 1$. The inverse of the coherence length $L$
for $S_1$ at $k=0$ versus $\delta$ is given in Fig. \ref{figJ12d_0_CL_k0pi}, where we can see that
as $\delta$ approaches 1, $L$ diverges. 
Because $\lambda=0$ is a critical point, we cannot get the series directly 
for the energy gap of $S_1$ at $k=0$. We can only get numerical
results for it by solving the integral equation\cite{longpaper}. This makes it poorly
convergent as $\lambda\to 1$, as we will see in the next section.
In this figure, we also plot the coherence length  $L$ at $k=\pi/2$ for $S_i$ and $T_i$ ($i=1,2$).

\begin{figure}
 \begin{center} 
 \vskip -0.5cm
 \epsfig{file=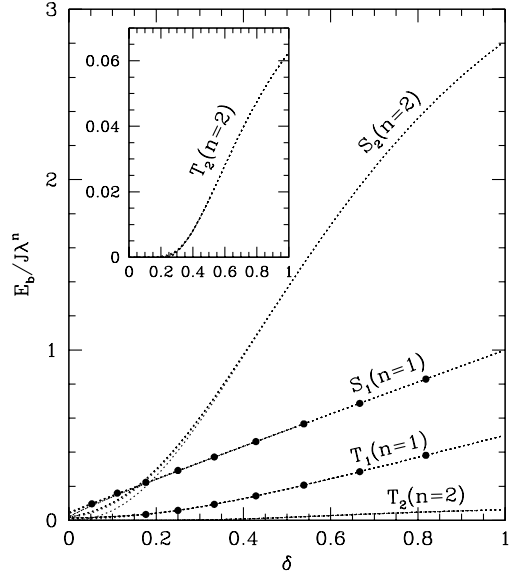, height=10cm}
 \vskip -1.2cm
 \caption[]
          {The scaled binding energy $E_b/J\lambda^n$ at $k=\pi/2$ versus dimerization 
          $\delta$ for two singlet ($S_1$ and $S_2$) and two triplet ($T_1$ and $T_2$)
         bound states of the $J_1-J_2-\delta$ chain with $\alpha=0$. The solid points are
         the numerical exact diagonalization results\cite{Barnes:99}.
         The inset enlarges the region  
         for $T_2$.
         Several different integrated differential approximants to
the series are shown.}
 \label{figJ12d_0_Eb_kpi} 
 \end{center}
\end{figure}

In the limit $\lambda\to 0$ ($\delta\to 1$), the binding energy at $k=\pi/2$ for 
$S_1$ and $T_1$ is proportional to $\lambda$, as expected, while for $S_2$ and $T_2$, the binding energy
is proportional to $\lambda^2$.
The rescaled binding energies $E_b/J\lambda^i$ versus $\delta$ for $S_i$ and $T_i$ ($i=1,2$) are shown in
Fig. \ref{figJ12d_0_Eb_kpi}. We also show some numerical exact 
diagonalization results\cite{Barnes:99} in this figure, which are in very good agreement
with our series results. 

As  evident from Figs. \ref{figJ12d_0_CL_k0pi} and \ref{figJ12d_0_Eb_kpi}  the bound state $T_2$
may disappear at about  $\delta=0.2$.

\subsubsection{Case $\alpha=(1- \delta)/2$}
Along the special
line $\alpha=(1- \delta)/2$, the ground state is an exact  product state, with 
the spins on each strongly-coupled  bond forming a singlet.
For non-zero $\delta$, the elementary
excitations for this system are triplets.
When $\delta \to 0$, the system has two degenerate ground states
and the triplets unbind into a pair of free spin-half
excitations. These spin-half objects, which are domain walls between the
two ground states, are called solitons and they become the elementary
excitations of the system.

The series for the irreducible two-particle matrix element $\Delta_2$ has 
been computed up to order $\lambda^{19}$ for 2-particle singlet, triplet
and quintet states \cite{wwwsite,footnote1} by using both orthogonal transformation 
(two block method) and similarity transformation methods\cite{longpaper}. Hence
one can  compute the series directly for the dispersion of the
bound states using a degenerate perturbation expansion. It turns out that
both transformations give identical series for the dispersion of the
bound states, up to the order computed,
 although the series for the irreducible two-particle matrix elements $\Delta_2$ are
different (the basis states are different in the two methods).
The energy gap at $k=\pi/2$ for one of the singlet bound
states,  $S_1$, is $1+3 \delta $ exactly\cite{cas82,sor99},
The series for the energy gaps of the other bound states
 (and also the lower edge of the continuum) at  $k=0$ and $\pi/2$ are given in Table \ref{tabJ12d_p5}.  

 \begin{figure}
 \begin{center} 
 \vskip -0.5cm
 \epsfig{file=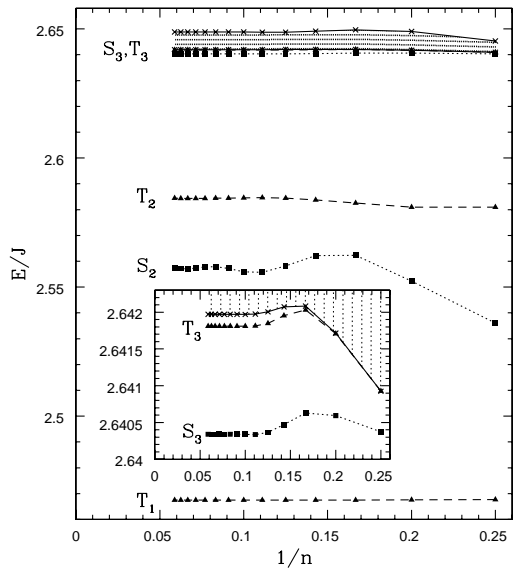, height=10cm}
 \vskip -1.2cm
 \caption[]
         {The energy gap $E/J$ at $k=\pi/2$ for two singlet ($S_2$ and $S_3$) 
         and three triplet ($T_1$, $T_2$ and $T_3$) bound states
           versus the inverse of order $1/n$ for the $J_1-J_2-\delta$ chain with
           $\delta=0.4$ and $\alpha=(1-\delta)/2$.
           The gray shaded regions are the two-particle continuum. 
           The inset enlarges the region near 
          $k=\pi/2$ so we can see $S_3$ and $T_3$ below the continuum.}
 \label{figJ12d_p5_E_ord_dp4} 
 \end{center}
\end{figure}

Here we find three singlet ($S_1$, $S_2$ and $S_3$) and three triplet, 
($T_1$, $T_2$ and $T_3$) 
bound states below the two-particle continuum, and two quintet antibound states
($Q_1$ and $Q_2$) above the continuum. 
The dispersions for these bound states at $\delta=0.4$ have been shown in Fig.4
of a preceding Letter \cite{shortpaper}.
To demonstrate the reliability of our results, we plot in Fig. 
\ref{figJ12d_p5_E_ord_dp4} the energy gap at $\delta=0.4$ and $k=\pi/2$
for all bound states and the two-particle continuum versus the inverse of the 
order $n$ up to $n=19$.
The results for $S_1$ are not plotted, since this case is known exactly. From this figure, we can see that the results are very well converged for $n>10$. 
In the limit of $\lambda\to 0$, the binding energy at $k=\pi/2$ for 
$S_1$ and $T_1$ is proportional to $\lambda$, as expected, while for $S_2$, $T_2$,
$S_3$ and $T_3$, the binding energy
is proportional to $\lambda^2$, $\lambda^2$, $\lambda^4$, $\lambda^6$
respectively, as we can see from Table \ref{tabJ12d_p5}.

\begin{figure}
 \begin{center} 
 \vskip -0.5cm
 \epsfig{file=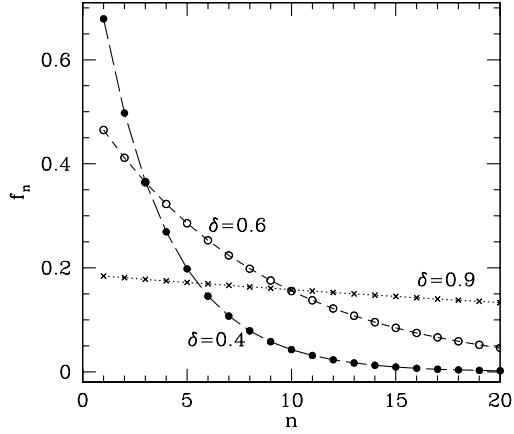, height=10cm}
 \vskip -1.2cm
 \caption[]
          {The amplitude $f_n$ versus $n$ for $T_3$ with $\delta=0.4$, 0.6 and 0.9, 
          and $k=\pi/2$}
 \label{figJ12d_p5_wf_T3} 
 \end{center}
\end{figure}

In the limit $\lambda\to 0$, the wave functions for $S_i$ and $T_i$ ($i=1,2$) 
are trivial, consisting of two triplets separated by $i-1$ singlet dimers, 
while the wave functions at $k=\pi/2$ for $S_3$ and $T_3$ are 
(here again we just take $S_z=0$ for $T_3$ as example)
\bea
\vert \psi_{S_3} (k) \rangle = && {1\over \sqrt{3N}} \sum_{j}  \sum_{n=1}^{\infty} e^{2ik(j+n+1/2)} f_n 
[t_{1}^{\dag} (j) t_{-1}^{\dag} (j+2n+1) \nonumber \\
&&  
+ t_{-1}^{\dag} (j) t_{1}^{\dag} (j+2n+1)
- t_{0}^{\dag} (j) t_{0}^{\dag} (j+2n+1) ] \vert 0 > \;, \\
\vert \psi_{T_3} (k) \rangle = && {1\over \sqrt{2N}} \sum_{j} \sum_{n=1}^{\infty} e^{2ik(j+n+1/2)} f_n
 [ t_{1}^{\dag} (j) t_{-1}^{\dag} (j+2n+1) 
- t_{-1}^{\dag} (j) t_{1}^{\dag} (j+2n+1) ] \vert 0 > \;.
\eea
For $S_3$, the amplitude $f_n$ for two triplets sitting at $j$ and $j+2n+1$ is
\be
f_n =  2^{-n} \sqrt{3} \;,
\ee
which decreases by a factor of 2 as $n$ increased by 1. We cannot obtain an 
analytic expression for $f_n$
for $T_3$, but the numerical results for $\delta=0.4$, 0.6 and 0.9 are presented in 
Fig. \ref{figJ12d_p5_wf_T3}:  for small $\lambda$ (large $\delta$), $f_n$ is almost independent of $n$,
so one has an infinite coherence length.
From the above results, one can see that the bound states $S_3$ and $T_3$ are
 due to the effective 
attraction between triplets separated by an even number (exclude 0) of singlet dimers.
Thus we find in this case that $S_i$ and $T_i$ ($i=1,2$) are fully localized states, 
whereas $S_3$ and $T_3$ are ``extended", with exponential tails to their wave functions.
As for the case $\alpha=0$, our calculations again show
 that for $k=\pi/2$ and any $\lambda$, the wavefunctions for $S_2$ and $T_2$ 
 only involve two triplets separated by an odd number of singlet dimers,
 while wavefunctions for $S_1$,  $T_1$, $S_3$, and $T_3$ only involve 
 two triplets separated by an odd number of singlet dimers.

\begin{figure}
 \begin{center} 
 \vskip -0.5cm
 \epsfig{file=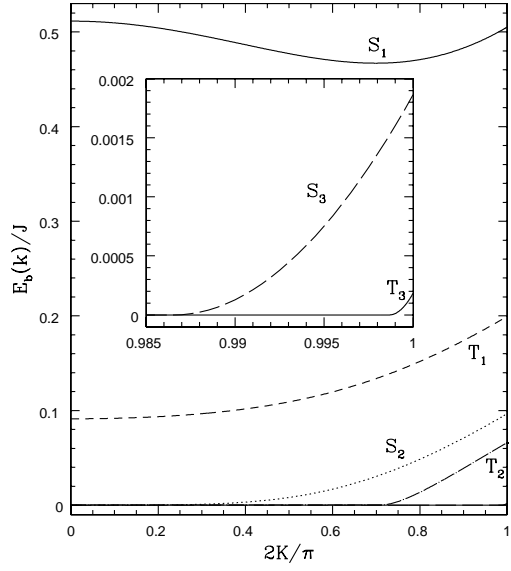, height=10cm}
 \vskip -1.2cm
 \caption[]
         {The binding energy $E_b$ for three singlet ($S_1$, $S_2$ and $S_3$) 
         and three triplet ($T_1$, $T_2$ and $T_3$) bound states
           versus momentum $k$ for the $J_1-J_2-\delta$ chain with
           $\delta=0.4$ and $\alpha=(1-\delta)/2$.
           The inset enlarges the region near $k=\pi/2$
            so one can see the nonzero binding energy for $S_3$, $T_3$.}
 \label{figJ12d_p5_Eb} 
 \end{center}
\end{figure}

\begin{figure}
 \begin{center} 
 \vskip -0.5cm
 \epsfig{file=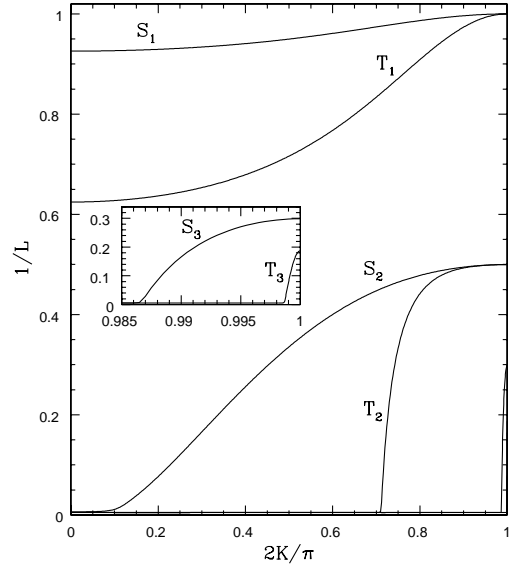, height=10cm}
 \vskip -1.2cm
 \caption[]
          {The inverse of the coherence length $1/L$ 
          versus momentum $k$ for three singlet ($S_1$, $S_2$ and $S_3$) 
         and three triplet ($T_1$, $T_2$ and $T_3$) 
          bound states of the $J_1-J_2-\delta$ chain 
          with $\delta=0.4$ and $\alpha=(1-\delta)/2$. The inset enlarges the region near 
          $k=\pi/2$.}
 \label{figJ12d_p5_CL_dp4} 
 \end{center}
\end{figure}

\begin{figure}
 \begin{center} 
 \vskip -0.5cm
 \epsfig{file=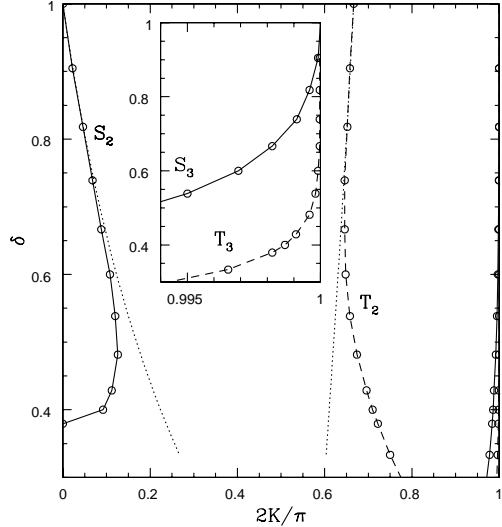, height=10cm}
  \vskip -1.2cm
 \caption[]
         {The critical $\delta$ versus momentum $k$ for singlet and triplet 
         bound states of the $J_1-J_2-\delta$ chain with $\alpha=(1-\delta)/2$. 
         The dotted lines are the
         results of Eq. (\ref{eqJ12dp5Kc}). }
 \label{figJ12d_p5_pd} 
 \end{center}   
\end{figure}

The two-particle binding energies and the inverse of the coherence 
lengths $1/L$ for $\delta=0.4$ are shown 
in Figs. \ref{figJ12d_p5_Eb} and \ref{figJ12d_p5_CL_dp4}. 
One can see that the singlet bound state $S_1$ and the triplet bound state $T_1$
exist for the whole range 
of momenta, while other
bound states exist only in a limited range of momenta $k>k_c$. 
 The results for
the critical momenta $k$ versus $\delta$ 
are given in Fig. \ref{figJ12d_p5_pd}, where one can see that 
in the limit $\delta\to 1$, the $S_3$ and $T_3$ states exist only at $k=\pi/2$.
For $S_2$ and $T_2$ bound states,
as before, one can  
get the first few terms in the series expansion for $k_c$:
\be
2k_c = \left\{ \begin{array}{ll}  
 2^{1/2} \lambda + 0.53033  \lambda^2 + O(\lambda^3) & \mbox{for $S_2$} \\
2 \pi/3 - 7 \lambda /(8\sqrt{3})  + 287 \lambda^2/(768\sqrt{3})+ O(\lambda^3)  & \mbox{for $T_2$}
\end{array}
\right.
\label{eqJ12dp5Kc}
\ee
and in the limit $k\to k_c$, the behaviour of the binding energy is
\bea
E_b/J\lambda^2 &=& 4 (k-k_c)^2 [ \lambda^2+ O(\lambda^3) ]/32  \nonumber \\
&& + 8 {(k-k_c)^3}\,
    \left( {\frac{\lambda}{32\,{\sqrt{2}}}} + 0.107723\,{\lambda^2} + O(\lambda^3) \right)
    +  O[(k-k_c)^4]
\eea
for $S_2$, and
\bea
E_b/J\lambda^2 &=&  
 4 {(k-k_c)^2}\,\left( {\frac{3}{32}} + {\frac{97\,\lambda}{512}} +
      {\frac{461\,{{\lambda}^2}}{1024}} + O(\lambda^3) \right)  \nonumber \\
      && +  8 {(k-k_c)^3}\,\left( {\frac{{\sqrt{3}}}{64}} -
      {\frac{199\,\lambda}{1024\,{\sqrt{3}}}} -
      {\frac{19273\,{{\lambda}^2}}{24576\,{\sqrt{3}}}} + O(\lambda^3) \right)
  + O[(k-k_c)^4]
\eea  
for $T_2$.
Here again one can see that  the ``critical index" is 2.
The results of Eq. (\ref{eqJ12dp5Kc}) are also shown in Fig. \ref{figJ12d_p5_pd}.
We can see that in the limit $\delta\to 1$, $k_c=0$ and $\pi/3$ for $S_2$ and $T_2$ respectively,
and as $\delta$ decreases, $k_c$ for $S_2$ firstly increases, then decreases back to
0 at around $\delta=0.38$, while $k_c$ for $T_2$
firstly decreases, then increases. We also can see from this figure that
$S_3$ and $T_3$ only exist over a tiny range of momenta for all $\delta$.

\begin{figure}
 \begin{center} 
 \vskip -0.5cm
 \epsfig{file=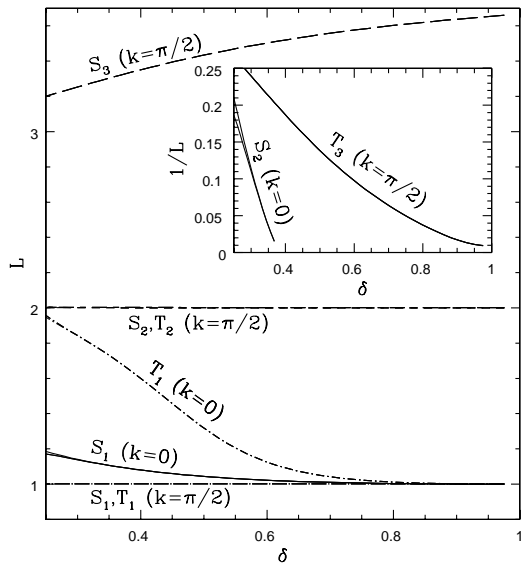, height=10cm}
 \vskip -1.2cm
 \caption[]
          {The  coherence length $L$ at $k=0$ and $\pi/2$
          versus $\delta$ for three singlet ($S_1$, $S_2$ and $S_3$) and
          three triplet ($T_1$, $T_2$ and $T_3$) 
          bound states of the $J_1-J_2-\delta$ chain 
          with $\alpha=(1-\delta)/2$. The inset plots $1/L$ versus $\delta$
          for $S_2$ at $k=0$ and $T_3$ at $k=\pi/2$. 
          The results of 17th to 19th order series are plotted. 
         }
 \label{figJ12d_p5_CL_k0pi} 
 \end{center}
\end{figure}

As for $S_1$ in the case $\alpha=0$, $\delta\to 1$ is also a critical point for $T_3$ at $k=\pi/2$.
The inverse of the coherence length $L$
for $T_3$ at $k=\pi/2$ and $S_2$ at $k=0$ versus $\delta$ 
are given in Fig. \ref{figJ12d_p5_CL_k0pi}, 
where we can see that
as $\delta$ approaches 1, the coherence length for $T_3$ at $k=\pi/2$ diverges,
while the bound state $S_2$ at $k=0$ appears at $\delta<0.38$, consistent with 
Fig. \ref{figJ12d_p5_pd}.
In this figure, we also plot the coherence length for other bound states
at $k=0$ and $\pi/2$. $L$ is exactly 1 for $S_1$ at $k=\pi/2$, while for $T_1$, $S_2$ and $T_2$
 at $k=\pi/2$, 
$L$ is almost equal to 1,2,2 respectively for all $\delta$. 
For $S_1$ and $T_1$ at $k=0$, $L$ is 1 in the limit $\delta\to 1$, and as
$\delta$ decreases, $L$ increases. 
For $S_3$ at $k=\pi/2$, $L$ is 11/3 in the limit $\delta\to 1$,
and as $\delta$ decreases, $L$ decreases. 

\section{ Regime of vanishing bond-alternation:
Unbinding of spin-half excitations}

In this section we turn to the regime of small $\delta$ or $\lambda$ near
unity. In this case, our results are less accurate and we have to rely
on series extrapolation methods. The limit $\delta\to 0$ is a critical
point, where we expect singularities in various physical quantities.
Hence, the convergence of the series breaks down and a simple truncation
does not lead to meaningful results. In this case,
we use the Dlog Pad\'e and  integrated differential approximants\cite{gut} to extrapolate
the series for the single-particle energies and the two-particle
binding energies. We present results based on these extrapolations.

We begin this section by making a few comments about the extrapolation
of one and two-particle energies to the $\delta\to 0$ limit for 
the nearest neighbor model ($\alpha=0$).
In this case the uniform system  at $\delta=0$
is the Bethe-ansatz solvable nearest-neighbor
Heisenberg model, with no gap in the excitation spectrum. 
Furthermore, it is believed that the mapping on to
the massive Thirring model gives the exact spectrum at small $k$
for small $\delta$ \cite{Uhrig:96}. 
The latter model has a well defined singlet and
a well defined triplet excitation whose energies are in the ratio of
$\sqrt{3}$. There are no further bound states so that there are no
triplet states between the single particle gap $\Delta$ and the two-particle
continuum gap $2\Delta$.

\begin{figure}
 \begin{center} 
 \vskip -0.5cm
 \epsfig{file=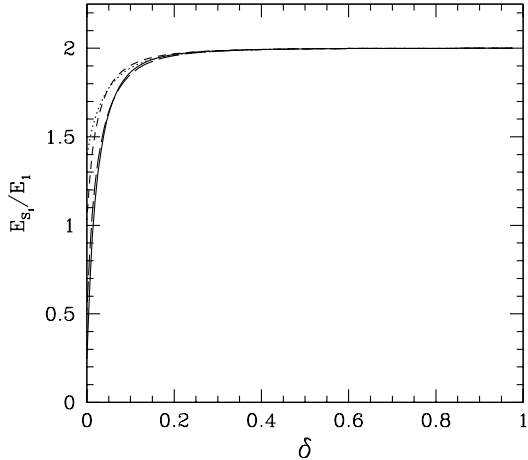, height=10cm}
 \vskip -1.2cm
 \caption[]
          {Ratio of the lowest singlet excitation energy $E_{S1}$ to the
lowest triplet excitation $E_1$, in the nearest neighbor model
as a function of $\delta$.
         Results at order 4 (dotted line), 5 (short dashed line), 6 (long dashed line) and
          7 (solid line)  are plotted.}
 \label{figJ12d_0_Eratio} 
 \end{center}
\end{figure}

As evident from figures
 \ref{figJ12d_0_CL_dp6},
 \ref{figJ12d_0_pd},  and
 \ref{figJ12d_0_CL_k0pi} only one singlet bound state exists as $\delta$
goes to zero for $k=0$. Its coherence length appears large but finite
in our calculations, though
its binding energy goes to zero. This, together with the elementary triplet
whose energy also goes to zero as $\delta$ goes to zero, gives
the two well defined states expected from the massive Thirring model.
As discussed in the previous section, the convergence of the singlet
excitation energy at $k=0$ becomes poor as $\lambda=1$, as it has to be
gotten from a numerical solution of the integral equation rather
than from an extrapolation of a binding energy series.
This ratio is plotted in Fig.
 \ref{figJ12d_0_Eratio}. It clearly stays close to $2$ down to
$\delta=0.2$ and begins decreasing for smaller values of $\delta$.
However, for $\delta<0.1$ the extrapolations become completely unreliable
and we cannot tell if the ratio approaches $\sqrt{3}$ as $\delta\to 0$. 

\begin{figure}
\begin{center} 
\vskip -0.5cm
\epsfig{file=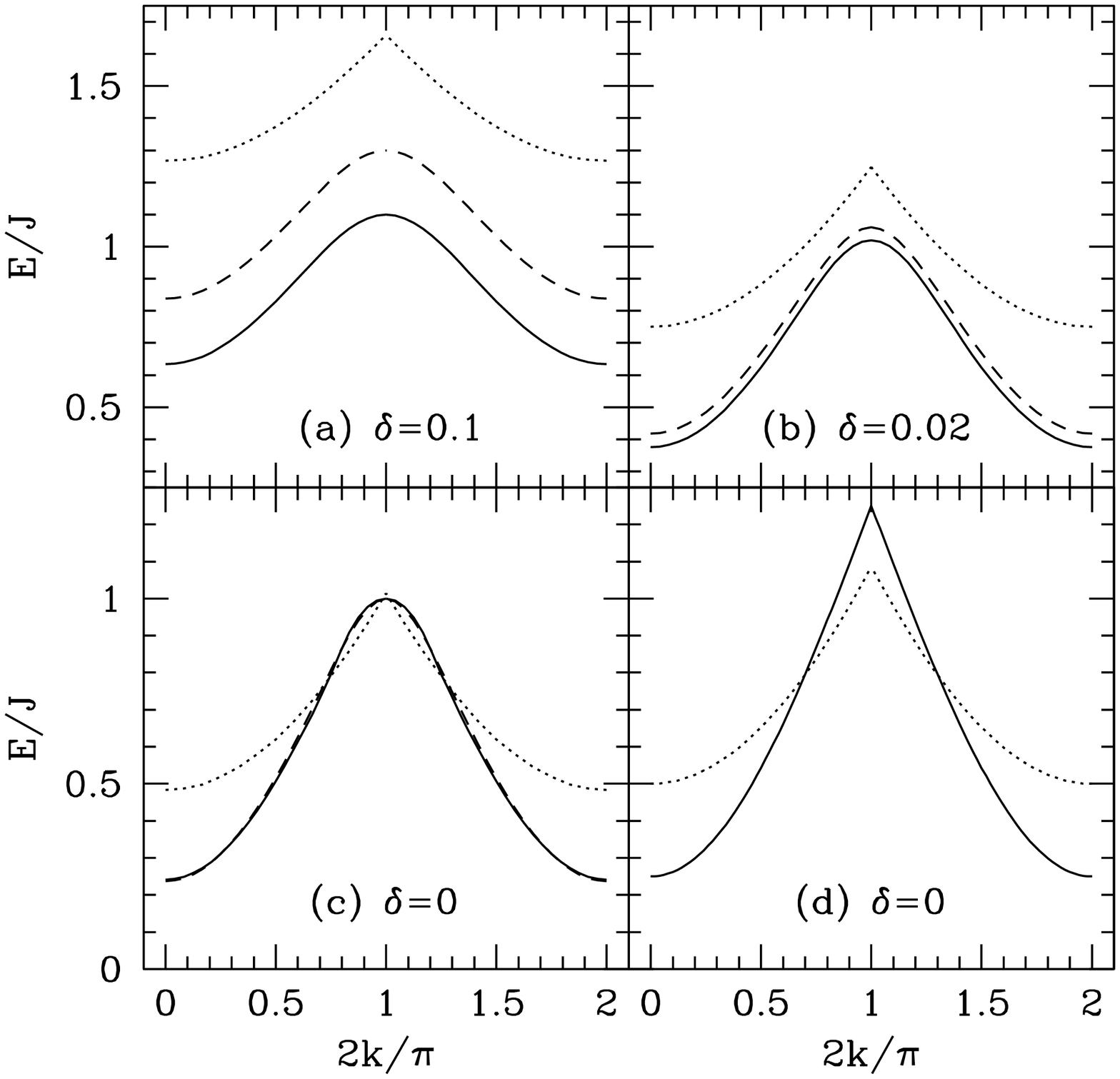, height=15cm}
\vskip -1.2cm
 \caption[]
          {Dispersion of the single-particle triplet excitation $E_1$ (solid line), 
          the lowest two-particle singlet bound state $S_1$ (dashed line), and the 
          bottom of the two-triplet continuua (dotted line)
          for  $\delta=0.1$ (a), 0.02 (b) and 0 (c) for the Shastry-Sutherland model. 
          Window (d) gives the
         variational results of Shastry and Sutherland, where only the
continuua are shown.
}
 \label{figJ12d_mk_E1S1} 
 \end{center}
\end{figure}

We now turn to the frustrated model
with $\alpha=(1- \delta)/2$. First, we show the dispersions of the 
single-particle triplet excitation, the lowest-energy
two-particle singlet bound state $S_1$, and the 
bottom of the two-triplet continuum over the full Brillouin
zone for various values of $\delta$ in Fig. \ref{figJ12d_mk_E1S1}.
It is evident that the triplet and the singlet spectra become degenerate as $\delta$
goes to zero. This is  direct evidence for free spin-half excitations,
since a pair of free spin-half excitations will form singlet and 
triplet states of equal energy.
Note that previous series expansion
studies have shown \cite{Rajiv:99} that the spectral weight
associated with the triplets does not vanish over the entire
Brillouin zone. It remains finite in the vicinity of $k=\pi/2$
at $\delta=0$. This result is consistent with the variational calculations
of Shastry and Sutherland,\cite{Shastry:81} who had found a soliton-antisoliton bound state
near $k=\pi/2$. Shastry and Sutherland had also found that the bound states
were four-fold degenerate meaning that  there are degenerate singlet
and triplet bound states. Thus, our results are completely consistent
with their calculations. S\o rensen and collaborators\cite{sor99} have also given an exact
demonstration of triplet and singlet states with energies $(1+\delta) J$
and $(1+3\delta ) J$ at $k=\pi/2$ respectively, and thus a splitting of 
$2\delta J$ which vanishes as $\delta\to 0$. Our results fit this pattern 
precisely.

Note further from this figure that as $\delta$ goes to zero,
in the region not too far from $k=\pi/2$,
the two-triplet continuum falls below
the states $S_1$ and $E_1$ (where $E_1$ is the single-particle triplet excitation). 
This is also true for the variational
calculation of Shastry and Sutherland shown in Fig. \ref{figJ12d_mk_E1S1}(d), 
though, to our knowledge,
it has not been noted before. 
At $k=\pi/2$, stable singlet and triplet states are rigorously known
to exist. A plausible picture is that a stable 
state exists only at or very near this wavevector,
and even at this point its binding energy with respect to the
multiparticle continuum is extremely small. It is also likely that the
spectral weight of these states is only appreciable in this narrow region;
further from $k=\pi/2$, these states will be lost in the continuum, and their
spectral weight will be neglible, in agreement with the 
calculation of Singh and Zheng\cite{Rajiv:99}.

\begin{figure}
\begin{center} 
\vskip -0.5cm
\epsfig{file=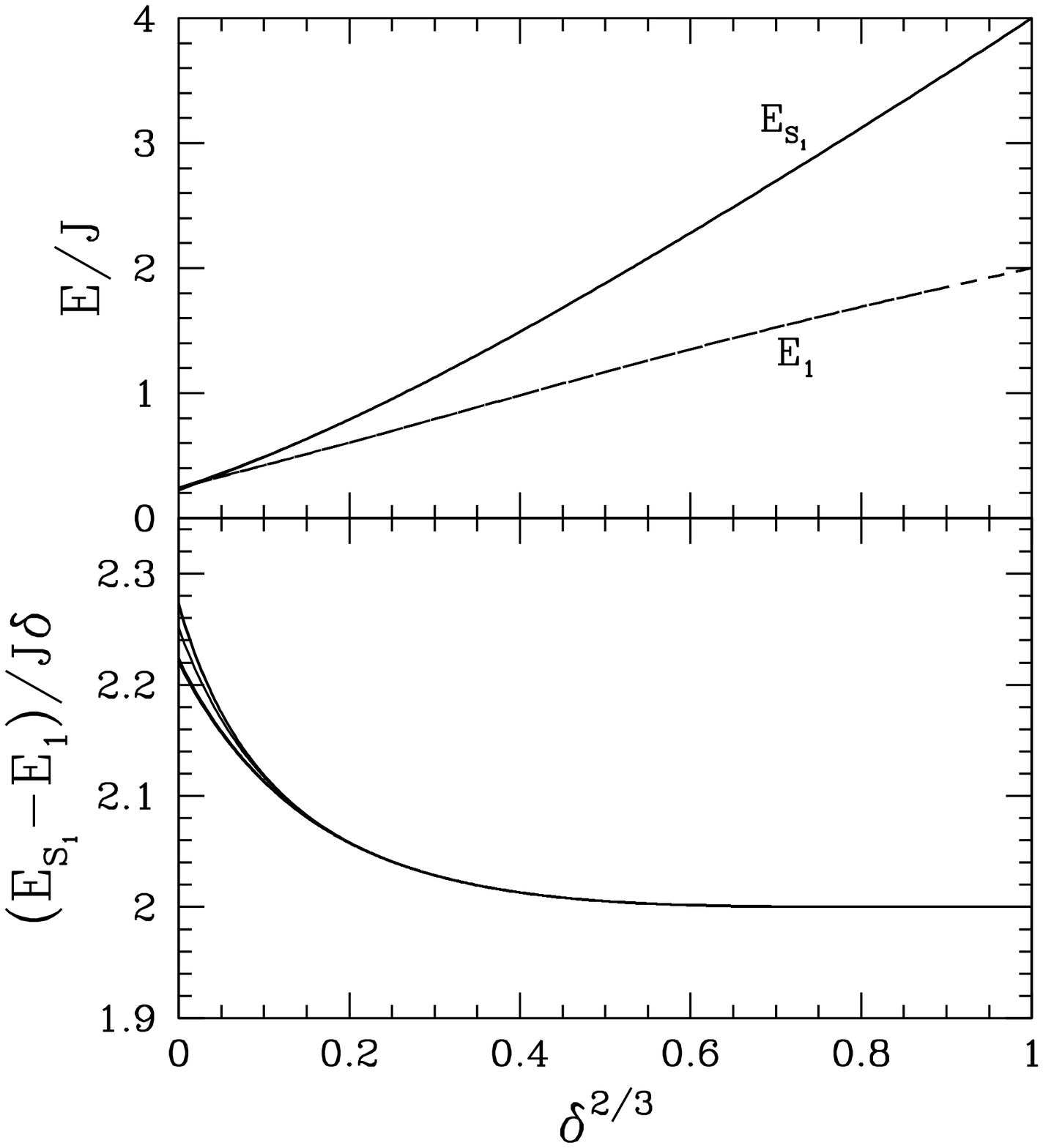, height=12cm}
\vskip -1.2cm
 \caption[]
          {The 1-particle triplet gap ($E_1$) and lowest singlet energy gap $E_{S_1}$ 
           at $k=0$ (upper window), and $(E_{S_1}-E_1)/J\delta $ (lower window) versus 
          $\delta^{2/3}$ 
          for the $J_1-J_2-\delta$ chain with $\alpha=(1-\delta)/2$. 
          The results of several different integrated differential 
         approximants to the series are shown.
}
 \label{figJ12d_E1_S1_k0} 
 \end{center}
\end{figure}

In the series extrapolation for the figure above, we have made use of the fact that  
as $\delta$ goes to zero (or $\lambda$ goes to 1), the energy gap approaches a
constant with correction proportional to $(1-\lambda)^{2/3}$\cite{oleg99}, so in the
series extrapolation we transform the series to a new variable
\begin{equation}
\lambda' = 1 - (1 - \lambda)^{2/3}
\end{equation}
to remove the singularity at $\lambda=1$.
Byrnes {\it et al.}\cite{oleg99} also predict that the singlet-triplet splitting
in the limit $\delta\to 0$ is 
\begin{equation}
(E_{S_1} - E_1)/J = c_1 \delta + c_2 \delta^{5/3} \label{eqSplitting} \;.
\end{equation}
Our series analysis seem to favor these arguments, as can be seen from Fig. \ref{figJ12d_E1_S1_k0}.
But our results give $c_1=2.25(5)$, rather than $15/8$ as given by  Byrnes {\it et al.}\cite{oleg99};
this difference is probably because we choose $\alpha=(1-\delta)/2$, while in
the calculations by Byrnes {\it et al.} $\alpha$ is fixed to be 1/2.

\begin{figure}
\begin{center} 
\vskip -0.5cm
\epsfig{file=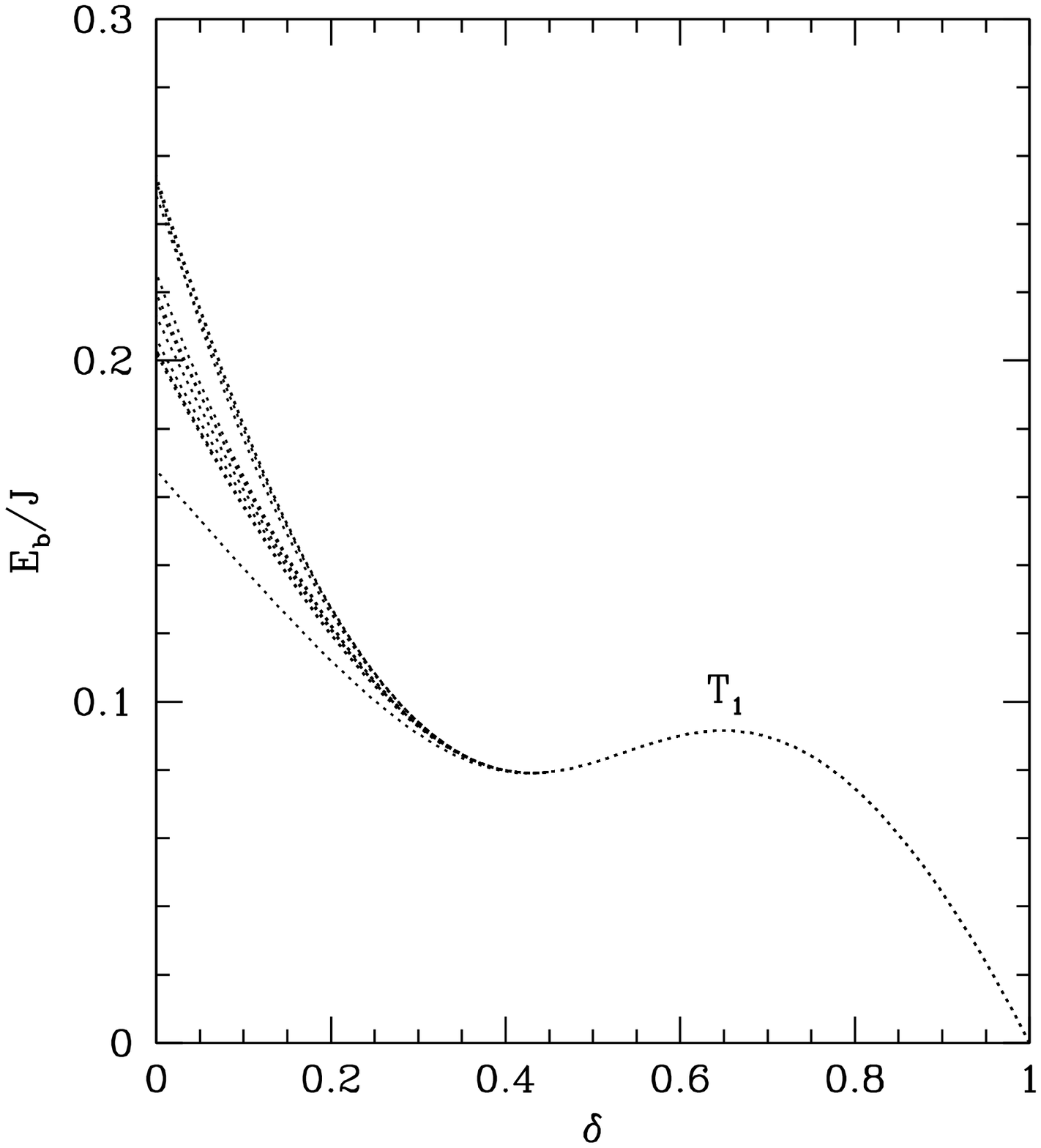, height=10cm}
\vskip -1.2cm
 \caption[]
          {The binding energy $E_b/J$ of the lowest triplet bound state ($T_1$) 
          at $k=0$ as a function of $\delta$.
           The results of several different integrated differential 
         approximants to the series are shown. }
 \label{e1t1s1}
 \end{center}  
\end{figure}

In Fig.~\ref{e1t1s1} the binding energy for the lowest triplet bound state 
$T_1$ is shown. It is clear that the extrapolation for $T_1$ does not converge 
very well. Looking at the figure, it 
is plausible to suggest that as $\delta$ goes to zero
the triplet binding energy also approaches $\Delta\approx 0.24J$, the single
particle energy gap. If true, this implies that this state also
becomes degenerate with the single particle state in this limit.
This is consistent with the idea that there are an infinite number
of states between the single particle gap $\Delta$ and the two-particle
continuum, which begins at $2\Delta$.
Thus the lowest of these states must approach $\Delta$ in
energy. However, we do not see a large number of two-particle bound states
in our calculations.
We suspect that the other bound states may arise from $3$, $4$, $5$, $\ldots$,
particle states. We hope to extend our methods to
study such multiparticle bound states in the future.

\begin{figure}
 \begin{center} 
\vskip -0.5cm
\epsfig{file=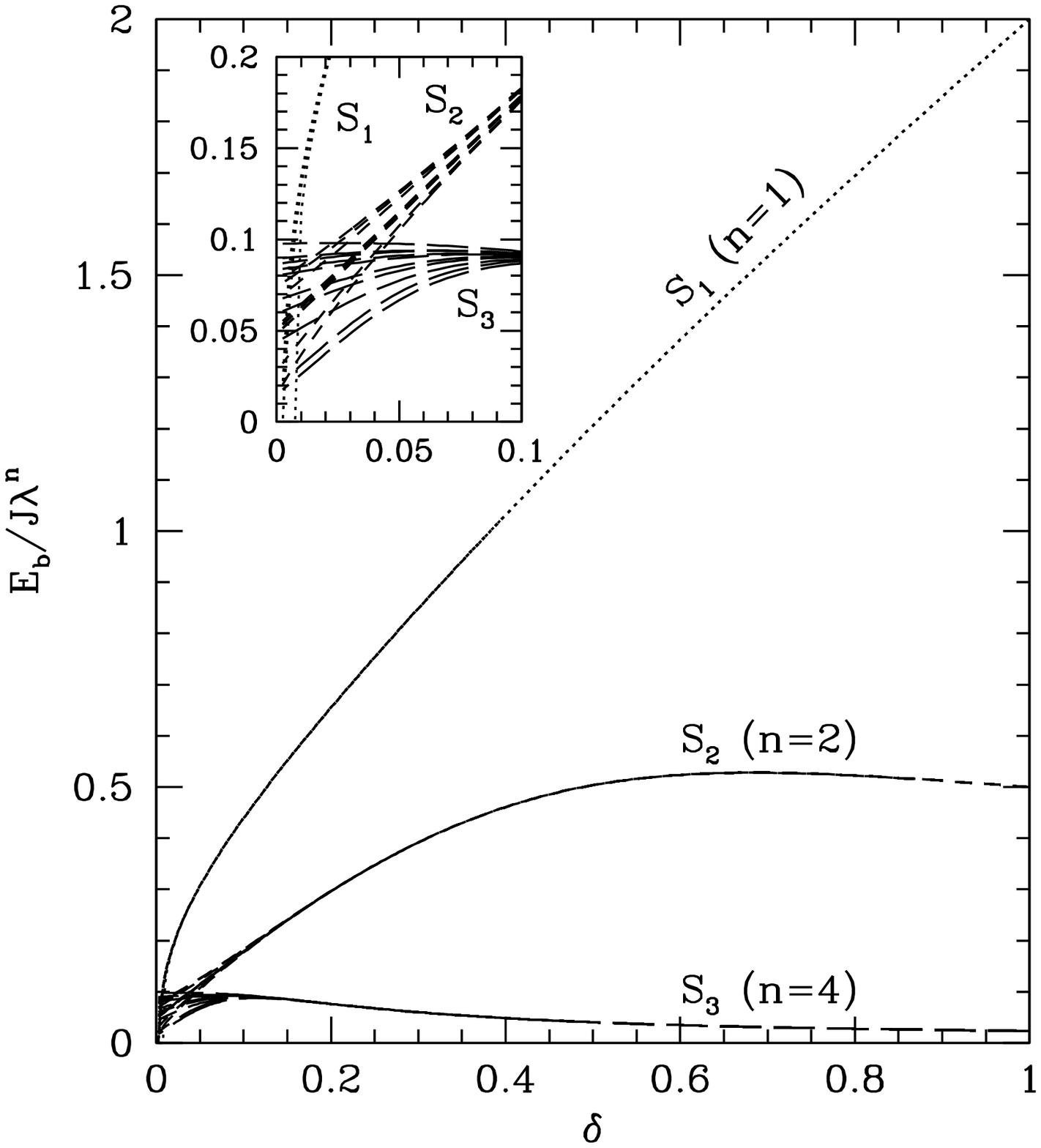, height=10cm}
\vskip -1.2cm
 \caption[]
          {The rescaled binding  energy $E_b/J\lambda^n$ at $k=\pi/2$ versus dimerization 
          $\delta$ for three singlet bound states $S_i$ ($i=1,2,3$)
          of the $J_1-J_2-\delta$ chain with $\alpha=(1-\delta)/2$.  The inset enlarges 
          the region near $\delta=0$.
         Several different integrated differential approximants to
   the series are shown.}
 \label{figJ12d_p5_Eb_Kpi_S123} 
 \end{center}
\end{figure}

\begin{figure}
 \begin{center} 
\vskip -0.5cm
\epsfig{file=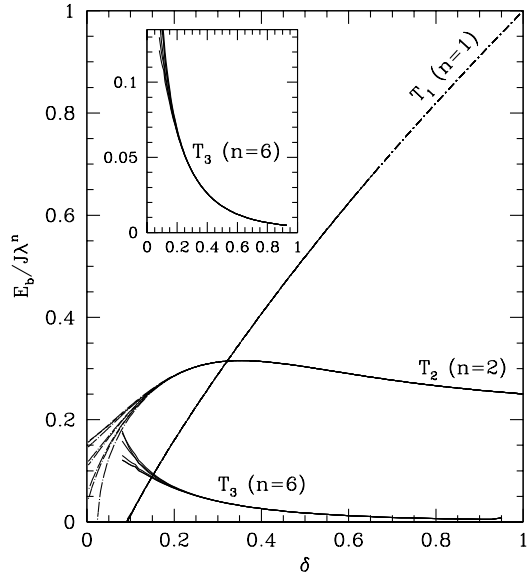, height=10cm}
\vskip -1.2cm
 \caption[]
          {The rescaled binding energy $E_b/J\lambda^n$ at $k=\pi/2$ versus dimerization 
          $\delta$ for three triplet bound states $T_i$ ($i=1,2,3$)
          of the $J_1-J_2-\delta$ chain with $\alpha=(1-\delta)/2$. The inset enlarges 
          the region for $T_3$.
         For $T_1$ and $T_2$, the results of several different integrated differential 
         approximants to the series are shown, while for $T_3$ the numerical results 
         from the integral equation 
         at order 14 to 19 are shown.}
 \label{figJ12d_p5_Eb_T123} 
 \end{center}
\end{figure}

We now turn to the bound states at $k=\pi/2$.
We show the rescaled singlet and triplet binding energies at $k=\pi/2$
in Figs. \ref{figJ12d_p5_Eb_Kpi_S123} and \ref{figJ12d_p5_Eb_T123}. 
These results for $S_i$ ($i=1,2,3$) and $T_i$ ($i=1,2$) are obtained from the
integrated differential approximants to the series, while the results for $T_3$ are
obtained from the numerical solution of the integral equation\cite{longpaper} at orders 14 to 19, 
since we cannot get the series
directly for $T_3$. We can see from these figures that as $\delta\to 0$, 
the binding energies for all $S_i$ ($i=1,2,3$)
approach the same value,  which is close to or equal to zero.
 Among $T_i$ ($i=1,2,3$),
$T_1$ has the largest binding energy in the limit $\delta\to 1$, but at $\delta=0.092(3)$,
the binding energy for $T_1$ becomes zero, while the binding
energies for $T_2$ and $T_3$ are still nonzero,
i.e., there appears to be a level crossing between $T_i$ ($i=1,2,3$) here: the
level crossing  between $T_1$ and $T_2$ happens at
$\delta=0.221(1)$ where the binding energy is $E_b/J=0.11955(10)$. 
The reason that  $T_1$ and $T_2$  can cross smoothly is presumably that
 the bound state for $T_1$ only 
involves two triplets separated by an even number of singlet dimers,
while the bound state for $T_2$ only 
involves two triplets separated by an odd number of singlet dimers, so that
the two states lie in disjoint sectors. 

\section{Conclusion and Discussion}

In this paper we have carried out an extensive investigation of the
two-particle spectra of the frustrated alternating Heisenberg chains
using strong coupling series expansions. 

In the regime of weakly coupled dimers,
the elementary excitations of the system
are triplets and the spin-half excitations are confined.
In this regime our series
expansions are convergent, and we have studied in great detail the
properties of the two-particle spectra, including binding energies,
coherence lengths and critical properties associated with vanishing
binding energies and diverging coherence lengths.
We find in every case, just as for the spin ladder system \cite{longpaper},
that where a 2-particle bound state emerges below the continuum at a 
``critical momentum" $k_c$, the binding energy behaves like $(k-k_c)^2$
as $k\to k_{c+}$, and the coherence length diverges, as one would expect.
Several distinct bound states can be identified, particularly near
$k=\pi/2$. Many of the bound states can only be seen by going to sufficiently
high orders in the perturbation expansion, showing the extended character
of the pair-attraction.

We have also studied the regime of deconfined spin-half excitations ($\delta\to 0$) by
using series extrapolation methods. Several properties of
one and two-particle spectra give a clear indication of this
deconfinement transition. 
The spectral weights of the triplets
vanish at the transition (except near $k=\pi/2$) and the singlet and triplet excitations
become degenerate. These methods can be used to look for such
unbinding transitions in higher dimensional models as well.

Our studies also raise several puzzles that need to be addressed
in the future. How does the spectrum between $\Delta$ and $2\Delta$
get filled up as one goes from the confined to the deconfined
phase? We suspect that multi-particle bound states, with varying
number of triplets are important. This needs to be further
addressed. 
The picture is much clearer in the soliton language, as one might expect:
at small $\delta$, there is a discrete spectrum of $s-\bar s$ bound states
confined by a linear potential, which becomes continuous as $\delta\to 0$ and the confining 
potential vanishes\cite{Affleck:97,Uhrig:99,oleg99}.
Another puzzle is the crossing of energy levels.
As $\lambda$ goes from zero to unity along the Shastry-Sutherland
line, one expects to see crossing of $n$-particle states, with
different $n$. How does this take place? How can this be
accounted for within the perturbation theory? Finally, are the
Shastry-Sutherland bound states stable away from $k=\pi/2$,
where they might decay into the $4$-soliton continuum?
We hope to address some of these issues in future. 

\acknowledgments

This work was initiated at the Quantum Magnetism program at the ITP at 
UC Santa Barbara which is supported by US National Science Foundation 
grant PHY94-07194.
The work of ZW and CJH was supported by a grant from the Australian
Research Council: they thank the New South Wales Centre for Parallel Computing 
for facilities and assistance with the calculations.
RRPS is supported in part by NSF grant number DMR-9986948.
ST gratefully acknowledges support by the German National Merit
Foundation and Bell Labs, Lucent Technologies.
HM wishes to thank the Yukawa Institute for Theoretical Physics for
hospitality. ZW and CJH would like to thank Prof. Oleg Sushkov
for some very useful discussions.

\newpage
\setdec 0.000000000000
\begin{table}
\squeezetable
\caption{Series coefficients for dimer expansions of the energy gap $E/J(1+\delta)$ 
 of two singlet bound states ($S_1$ and $S_2$), 
          two triplet bound states ($T_1$ and $T_2$), two
          quintet antibound states ($Q_1$ and $Q_2$), and the lower edge and upper edge of 
          the continuum ($C_l$ and $C_u$) at $k=\pi/2$ for the  $J_1-J_2-\delta$ chain with
           $\alpha=0$.
Nonzero coefficients $\lambda^n$
up to order $n=11$  are listed.}\label{tabJ12d_0}
\begin{tabular}{rrrrr}
\multicolumn{1}{c}{$n$} &\multicolumn{1}{c}{$E_{S_1}/J(1+\delta)$ for $S_1$} &\multicolumn{1}{c}{$E_{S_2}/J(1+\delta)$ for $S_2$}
&\multicolumn{1}{c}{$E_{T_1}/J(1+\delta)$ for $T_1$}  &\multicolumn{1}{c}{$E_{T_2}/J(1+\delta)$ for $T_2$} \\
\hline
  0 &\dec   2.000000000       &\dec   2.000000000       &\dec   2.000000000       &\dec   2.000000000       \\
  1 &\dec  -5.000000000$\times 10^{-1}$ &\dec   0.000000000       &\dec  -2.500000000$\times 10^{-1}$ &\dec   0.000000000       \\
  2 &\dec   1.875000000$\times 10^{-1}$ &\dec  -3.906250000$\times 10^{-1}$ &\dec   1.562500000$\times 10^{-1}$ &\dec  -2.812500000$\times 10^{-1}$ \\
  3 &\dec   3.906250000$\times 10^{-2}$ &\dec   7.812500000$\times 10^{-2}$ &\dec  -2.343750000$\times 10^{-2}$ &\dec   6.250000000$\times 10^{-2}$ \\
  4 &\dec  -2.050781250$\times 10^{-2}$ &\dec   3.324584961$\times 10^{-1}$ &\dec  -6.168619792$\times 10^{-2}$ &\dec   9.326171875$\times 10^{-2}$ \\
  5 &\dec  -4.154459635$\times 10^{-2}$ &\dec  -5.464426676$\times 10^{-2}$ &\dec  -8.189561632$\times 10^{-2}$ &\dec  -1.143256293$\times 10^{-1}$ \\
  6 &\dec  -4.004075792$\times 10^{-2}$ &\dec  -4.499045478$\times 10^{-1}$ &\dec  -9.442308214$\times 10^{-2}$ &\dec  -1.716450585$\times 10^{-1}$ \\
  7 &\dec  -2.573965214$\times 10^{-2}$ &\dec  -1.200470705$\times 10^{-1}$ &\dec  -8.740929615$\times 10^{-2}$ &\dec  -2.332303701$\times 10^{-2}$ \\
  8 &                         &                         &\dec  -6.322668408$\times 10^{-2}$ &\dec   2.681890517$\times 10^{-2}$ \\
  9 &                         &                         &\dec  -2.250317901$\times 10^{-2}$ &\dec  -3.523763254$\times 10^{-2}$ \\
 10 &                         &                         &\dec   2.941323970$\times 10^{-2}$ &\dec   1.450322827$\times 10^{-2}$ \\
 11 &                         &                         &\dec   9.059538250$\times 10^{-2}$ &\dec   1.467690427$\times 10^{-1}$ \\
\hline
\multicolumn{1}{c}{$n$} &\multicolumn{1}{c}{$E_{Q_2}/J(1+\delta)$ for $Q_2$} 
&\multicolumn{1}{c}{$E_{Q_1}/J(1+\delta)$ for $Q_1$}
&\multicolumn{1}{c}{$E_{C_l}/J(1+\delta)$ for $C_l$} &\multicolumn{1}{c}{$E_{C_u}/J(1+\delta)$ for $C_u$} \\
\hline
  0 &\dec   2.000000000       &\dec   2.000000000       &\dec   2.000000000       &\dec   2.000000000       \\
  1 &\dec   0.000000000       &\dec   2.500000000$\times 10^{-1}$ &\dec   0.000000000       &\dec   0.000000000       \\
  2 &\dec   3.125000000$\times 10^{-2}$ &\dec  -9.375000000$\times 10^{-2}$ &\dec  -2.500000000$\times 10^{-1}$ &\dec   0.000000000       \\
  3 &\dec   1.250000000$\times 10^{-1}$ &\dec  -1.015625000$\times 10^{-1}$ &\dec   3.125000000$\times 10^{-2}$ &\dec   1.562500000$\times 10^{-1}$ \\
  4 &\dec   1.513671875$\times 10^{-2}$ &\dec  -5.126953125$\times 10^{-2}$ &\dec  -2.343750000$\times 10^{-2}$ &\dec   9.375000000$\times 10^{-2}$ \\
  5 &\dec   1.916503906$\times 10^{-2}$ &\dec  -1.114908854$\times 10^{-2}$ &\dec  -6.868489583$\times 10^{-2}$ &\dec  -7.269965278$\times 10^{-3}$ \\
  6 &\dec   5.786874559$\times 10^{-2}$ &\dec  -6.629096137$\times 10^{-4}$ &\dec   4.757351345$\times 10^{-3}$ &\dec  -1.196628147$\times 10^{-2}$ \\
  7 &\dec   2.270891195$\times 10^{-2}$ &\dec   1.044698998$\times 10^{-2}$ &\dec  -1.730789373$\times 10^{-3}$ &\dec   8.809831407$\times 10^{-3}$ \\
  8 &\dec   1.947733239$\times 10^{-2}$ &\dec   2.007299524$\times 10^{-2}$ &\dec  -3.478056560$\times 10^{-2}$ &\dec   4.201953303$\times 10^{-3}$ \\
  9 &\dec   2.277670953$\times 10^{-2}$ &\dec   3.177196538$\times 10^{-2}$ &\dec   2.936064788$\times 10^{-4}$ &\dec  -2.457848151$\times 10^{-3}$ \\
 10 &\dec  -1.225328947$\times 10^{-3}$ &\dec   4.251490921$\times 10^{-2}$ &\dec   5.560815615$\times 10^{-3}$ &\dec   8.842688260$\times 10^{-4}$ \\
 11 &\dec  -9.435317910$\times 10^{-3}$ &\dec   5.454945223$\times 10^{-2}$ &\dec  -2.004222569$\times 10^{-2}$ &\dec   2.355556912$\times 10^{-3}$ \\
\end{tabular}
\end{table}

\setdec 0.000000000000
\begin{table}
\squeezetable
\caption{Series coefficients for dimer expansions of the energy gap $E/J(1+\delta)$ 
 of  singlet bound states ($S_1$, $S_2$, $S_3$), 
          triplet bound states ($T_1$ and $T_2$) and the lower edge of 
          the continuum ($C_l$) at $k=0$ and $\pi/2$ for the  $J_1-J_2-\delta$ chain with
           $\alpha=(1-\delta)/2$.
Nonzero coefficients of $\lambda^n$
up to order $n=19$  are listed.}\label{tabJ12d_p5}
\begin{tabular}{rrrrr}
\multicolumn{1}{c}{$n$} &\multicolumn{1}{c}{$S_1$ at $k=0$} 
&\multicolumn{1}{c}{$T_1$ at $k=0$}
&\multicolumn{1}{c}{$C_l$ at $k=0$} 
 &\multicolumn{1}{c}{$S_2$ at $k=\pi/2$} \\
\hline
  0 &\dec   2.000000000       &\dec   2.000000000       &\dec   2.000000000       &\dec   2.000000000       \\
  1 &\dec  -1.000000000       &\dec  -5.000000000$\times 10^{-1}$ &\dec   0.000000000       &\dec   0.000000000       \\
  2 &\dec  -5.000000000$\times 10^{-1}$ &\dec   1.250000000$\times 10^{-1}$ &\dec  -1.000000000       &\dec  -7.500000000$\times 10^{-1}$ \\
  3 &\dec  -2.500000000$\times 10^{-1}$ &\dec  -1.562500000$\times 10^{-1}$ &\dec  -5.000000000$\times 10^{-1}$ &\dec  -6.250000000$\times 10^{-1}$ \\
  4 &\dec   6.250000000$\times 10^{-2}$ &\dec  -9.257812500$\times 10^{-1}$ &\dec   6.250000000$\times 10^{-2}$ &\dec  -4.687500000$\times 10^{-2}$ \\
  5 &\dec   2.031250000$\times 10^{-1}$ &\dec  -1.833007813       &\dec   3.593750000$\times 10^{-1}$ &\dec   8.066406250$\times 10^{-1}$ \\
  6 &\dec   3.255208333$\times 10^{-2}$ &\dec  -2.977783203       &\dec   1.542968750$\times 10^{-1}$ &\dec   1.142822266       \\
  7 &\dec  -2.444118924$\times 10^{-1}$ &\dec  -4.028717041       &\dec  -2.992621528$\times 10^{-1}$ &\dec  -2.324761285$\times 10^{-2}$ \\
  8 &\dec  -2.273898655$\times 10^{-1}$ &\dec  -3.409357212       &\dec  -4.475063748$\times 10^{-1}$ &\dec  -2.524002923       \\
  9 &\dec   1.677377960$\times 10^{-1}$ &\dec   4.411956160       &\dec  -2.119700114$\times 10^{-2}$ &\dec  -3.662303224       \\
 10 &\dec   4.367628576$\times 10^{-1}$ &\dec   3.284248577$\times 10^{1}$ &\dec   5.406151312$\times 10^{-1}$ &\dec   6.600969298$\times 10^{-1}$ \\
 11 &\dec   4.447491734$\times 10^{-2}$ &\dec   1.064643399$\times 10^{2}$ &\dec   4.678734733$\times 10^{-1}$ &\dec   1.029881729$\times 10^{1}$ \\
 12 &\dec  -6.649650688$\times 10^{-1}$ &\dec   2.537572572$\times 10^{2}$ &\dec  -3.650978943$\times 10^{-1}$ &\dec   1.420401469$\times 10^{1}$ \\
 13 &\dec  -5.647530897$\times 10^{-1}$ &\dec   4.601292420$\times 10^{2}$ &\dec  -9.976553176$\times 10^{-1}$ &\dec  -5.430672245       \\
 14 &\dec   6.692916069$\times 10^{-1}$ &\dec   5.089128172$\times 10^{2}$ &\dec  -3.568290713$\times 10^{-1}$ &\dec  -4.753274080$\times 10^{1}$ \\
 15 &\dec   1.434455637       &\dec  -4.198436299$\times 10^{2}$ &\dec   1.149565251       &\dec  -6.050262051$\times 10^{1}$ \\
 16 &\dec  -7.932368314$\times 10^{-2}$ &\dec  -4.530966202$\times 10^{3}$ &\dec   1.548592323       &\dec   3.784212529$\times 10^{1}$ \\
 17 &\dec  -2.535956672       &\dec  -1.638065807$\times 10^{4}$ &\dec  -3.749497848$\times 10^{-1}$ &\dec   2.350096418$\times 10^{2}$ \\
 18 &\dec  -1.777648180       &\dec  -4.239290399$\times 10^{4}$ &\dec  -2.734332838       &\dec   2.710161551$\times 10^{2}$ \\
 19 &\dec   3.125549537       &\dec  -8.305280519$\times 10^{4}$ &\dec  -1.897635556       &\dec  -2.494277528$\times 10^{2}$ \\
\hline
\multicolumn{1}{c}{$n$} &\multicolumn{1}{c}{$S_3$ at $k=\pi/2$} 
&\multicolumn{1}{c}{$T_1$ at $k=\pi/2$}
&\multicolumn{1}{c}{$T_2$ at $k=\pi/2$} 
 &\multicolumn{1}{c}{$C_l$ at $k=\pi/2$} \\
\hline
  0 &\dec   2.000000000       &\dec   2.000000000       &\dec   2.000000000       &\dec   2.000000000       \\
  1 &\dec   0.000000000       &\dec  -5.000000000$\times 10^{-1}$ &\dec   0.000000000       &\dec   0.000000000       \\
  2 &\dec  -5.000000000$\times 10^{-1}$ &\dec  -1.250000000$\times 10^{-1}$ &\dec  -6.250000000$\times 10^{-1}$ &\dec  -5.000000000$\times 10^{-1}$ \\
  3 &\dec  -2.500000000$\times 10^{-1}$ &\dec   0.000000000       &\dec  -4.375000000$\times 10^{-1}$ &\dec  -2.500000000$\times 10^{-1}$ \\
  4 &\dec  -7.421875000$\times 10^{-2}$ &\dec  -3.906250000$\times 10^{-3}$ &\dec  -2.148437500$\times 10^{-1}$ &\dec  -6.250000000$\times 10^{-2}$ \\
  5 &\dec   1.123046875$\times 10^{-2}$ &\dec  -6.835937500$\times 10^{-3}$ &\dec   2.929687500$\times 10^{-3}$ &\dec   3.906250000$\times 10^{-2}$ \\
  6 &\dec   5.106608073$\times 10^{-3}$ &\dec  -5.777994792$\times 10^{-3}$ &\dec   1.789143880$\times 10^{-1}$ &\dec   4.231770833$\times 10^{-2}$ \\
  7 &\dec  -4.116566976$\times 10^{-2}$ &\dec  -8.443196615$\times 10^{-4}$ &\dec   3.243408203$\times 10^{-1}$ &\dec  -3.363715278$\times 10^{-3}$ \\
  8 &\dec  -6.972208729$\times 10^{-2}$ &\dec  -4.641285649$\times 10^{-3}$ &\dec   4.088701319$\times 10^{-1}$ &\dec  -4.214590567$\times 10^{-2}$ \\
  9 &\dec  -5.548241367$\times 10^{-2}$ &\dec   2.439375277$\times 10^{-4}$ &\dec   3.233009297$\times 10^{-1}$ &\dec  -4.348302771$\times 10^{-2}$ \\
 10 &\dec  -1.615796414$\times 10^{-2}$ &\dec  -5.090890892$\times 10^{-3}$ &\dec  -7.412812403$\times 10^{-2}$ &\dec  -1.641364313$\times 10^{-2}$ \\
 11 &\dec   1.393358333$\times 10^{-2}$ &\dec   2.287079761$\times 10^{-3}$ &\dec  -7.776608991$\times 10^{-1}$ &\dec   9.839517861$\times 10^{-3}$ \\
 12 &\dec   1.628315511$\times 10^{-2}$ &\dec  -6.726104563$\times 10^{-3}$ &\dec  -1.459665977       &\dec   1.402656002$\times 10^{-2}$ \\
 13 &\dec   1.092421645$\times 10^{-3}$ &\dec   4.895928063$\times 10^{-3}$ &\dec  -1.501800181       &\dec  -1.566731983$\times 10^{-3}$ \\
 14 &\dec  -7.495980128$\times 10^{-3}$ &\dec  -9.649045824$\times 10^{-3}$ &\dec  -3.338269791$\times 10^{-1}$ &\dec  -1.834857746$\times 10^{-2}$ \\
 15 &\dec   4.106539880$\times 10^{-3}$ &\dec   8.959512690$\times 10^{-3}$ &\dec   2.080412782       &\dec  -2.058150001$\times 10^{-2}$ \\
 16 &\dec   2.756737781$\times 10^{-2}$ &\dec  -1.466894655$\times 10^{-2}$ &\dec   4.898068211       &\dec  -8.451718109$\times 10^{-3}$ \\
 17 &\dec   4.306603014$\times 10^{-2}$ &\dec   1.550047623$\times 10^{-2}$ &\dec   6.361874618       &\dec   4.993840865$\times 10^{-3}$ \\
 18 &\dec   3.910407449$\times 10^{-2}$ &\dec  -2.303464950$\times 10^{-2}$ &\dec   4.129539836       &\dec   7.616721351$\times 10^{-3}$ \\
 19 &\dec   2.231265079$\times 10^{-2}$ &\dec   2.621842144$\times 10^{-2}$ &\dec  -3.760012157       &\dec  -1.055977079$\times 10^{-3}$ \\
\end{tabular}
\end{table}

\end{document}